\documentclass[aps,pre,twocolumn,superscriptaddress,showpacs]{revtex4}
\usepackage{amssymb}
\usepackage{epsfig}
\usepackage{amsmath}
\usepackage{times}

\begin{document}

\title{On the pinning strategy of complex networks}
\author{Lei Yang}\affiliation{Institute for Fusion Theory and Simulation, Zhejiang
University, Hangzhou, China 310027}
\author{Xingang Wang}
\email[Corresponding author. Email address: ]{wangxg@zju.edu.cn}
\affiliation{Institute for Fusion Theory and Simulation, Zhejiang
University, Hangzhou, China 310027} \affiliation{Department of
Physics, Zhejiang University, Hangzhou, China 310027}
\author{Yao Li}
\affiliation{Institute for Fusion Theory and Simulation, Zhejiang
University, Hangzhou, China 310027}
\author{Zhengmao Sheng}
\affiliation{Institute for Fusion Theory and Simulation, Zhejiang
University, Hangzhou, China 310027}

\begin{abstract}
In pinning control of complex networks, a tacit believing is that
the system dynamics will be better controlled by pinning the
large-degree nodes than the small-degree ones. Here, by changing the
number of pinned nodes, we find that, when a significant fraction of
the network nodes are pinned, pinning the small-degree nodes could
generally have a higher performance than pinning the large-degree
nodes. We demonstrate this interesting phenomenon on a variety of
complex networks, and analyze the underlying mechanisms by the model
of star networks. By changing the network properties, we also find
that, comparing to densely connected homogeneous networks, the
advantage of the small-degree pinning strategy is more distinct in
sparsely connected heterogenous networks.

\end{abstract}
\date{\today }
 \maketitle

Pinning control refers to the controlling of system dynamics to a
target state by perturbing only partial of the system variables
\cite{HQ:1994,HZH:1995}. Due to its convenient operation and high
performance, this technique has been widely used in controlling
spatiotemporal dynamics in complex systems, including optical
systems \cite{Appl:Optics}, plasma \cite{Appl:Plasma}, neural
systems \cite{Appl:Neoron}, and chemical and flow turbulence in
various systems
\cite{GCS:1997,Appl:turbulence3,Appl:turbulence1,Appl:turbulence2}.
In pinning control, one of the focusing issues is about how to
optimize the pinning scheme according to the system topology
\cite{GCS:1997,Appl:turbulence3}. In controlling the dynamics of
coupled map lattices, it is found that, with the same pinning
effort, the controllability can be significantly improved if the
pinnings are arranged in an asymmetric fashion \cite{GCS:1997}. The
underlying mechanism for this improvement is that, by asymmetric
pinning scheme, the topology of the lattice network is actually
modified from symmetric to asymmetric, which causes a more uniform
suppression of the unstable modes \cite{GCS:1997}. Still in lattice
network, based on the special property of coupling gradient, it is
even found that the spatiotemporal chaos can be successfully
controlled by pinning a single node \cite{HYXQ:1997}. In this
application, to identify the location of the pinned node, we need to
analyze the topology of the weighted and directed lattice
\cite{HYXQ:1997}.

Comparing to the regular networks, optimization of pinning scheme
according to topology properties is more significant in complex
networks \cite{XFW:2002,XL:2004,TC:2007,SF:2007,ZC:2008,ZLZ:2009}.
For practical networks, a general feature is that the number of
links (degree) associated to the network node has a large variation,
e.g. the type of scale-free networks (SFN) \cite{CN:REV}. In
networks like SFN, the majority of the network nodes have small
degree, but a few of the network nodes could have distinctively
large degree. Noticed the important roles of the large-degree nodes
played in other types of network dynamics, e.g. network
synchronization \cite{SYN:REV}, it is natural to expect that the
spatiotemporal chaos of complex networks will be better controlled
by pinning the large-degree nodes than pinning the small-degree ones
or pinning randomly. To check this, in Ref. \cite{XFW:2002} the
authors compared the performance of two pinning schemes on SFN:
large-degree pinning (LDP) and random pinning. In LDP, the nodes are
pinned by the decreasing order of their degree, while in random
pinning the nodes are randomly selected. The authors found that,
with the same number of pinned nodes, the LDP strategy has a clear
advantage over the random pinning strategy. After the work of Ref.
\cite{XFW:2002}, pinning control of complex networks has received
continuous interest, and the advantage of LDP has been well
addressed \cite{TC:2007,SF:2007,ZC:2008,ZLZ:2009}.

In a practical situation, due to the limited pinning strength, to
control a large-size complex network, a number of the network nodes
should be pinned simultaneously \cite{XFW:2002,XL:2004}. For
instance, with a reasonable pinning strength (even with the order of
the dynamics variables), it is shown that the number of pinned nodes
necessary for controlling a SFN to a homogeneous state could take up
to 30 or 50 percent of the total network nodes \cite{XFW:2002}. With
such a significant fraction of nodes be pinned, it is questionable
whether the LDP strategy will be still the best option, as now many
of the pinned nodes do not have large degree. In this paper, we will
re-evaluate the performance of LDP by comparing it with its opposite
case, namely the small-degree pinning (SDP) strategy. In contrast to
LDP, in SDP the network nodes are pinned by the increasing order of
their degree. Very interestingly, we found that when the fraction of
pinned nodes is large enough (depending on the network details, this
fraction could be ranging from 30 to 50 percent), the SDP strategy,
which has been tacitly believed as having the worst performance,
will outperform the LDP strategy. We demonstrate this phenomenon in
various complex networks, and investigate its underlying mechanism
by a simplified network model.

Our model of network pinning is the following
\begin{equation}
\dot{\mathbf{x}}_i = \mathbf{F}(\mathbf{x_i}) + \varepsilon
\sum_{j=1}^{N}a_{ij}
[\mathbf{H}(\mathbf{x_j})-\mathbf{H}(\mathbf{x_i})]+\eta\sum_{m\in
V}\delta_{im}[\mathbf{H}(\mathbf{x}_T)-\mathbf{H}(\mathbf{x}_i)],
\label{model}
\end{equation}
where $i,j=1,2,\ldots,N$ are the node indices, $\mathbf{F}$ is the
function of the node dynamics, $\mathbf{H}$ is the coupling
function. $\varepsilon$ and $\eta$ represent the uniform coupling
strength and pinning strength, respectively. The network structure
is captured by the adjacency matrix $A=\{a_{ij}\}$, in which
$a_{ij}=1$ if nodes $i$ and $j$ are directly connected, and
$a_{ij}=0$ otherwise. The degree of node $i$ thus reads
$k_i=\sum_{j=1}^{N}a_{ij}$. The target orbit to which the whole
network is assumed to be controlled is denoted by $\mathbf{x}_T$.
The total number of pinned nodes is $n$, and the set of pinned nodes
is represented by $V=\left\{ m \right\}$. Specifically, if node $i$
is pinned in the network, we have $\delta_{im}=1$ in Eq. [1],
otherwise $\delta_{im}=0$. Without loss of generality, we set the
target (controller) to be having the same dynamics as the network
node, i.e. $\dot{\mathbf{x}}_T=\mathbf{F}(\mathbf{x_T})$.

Regarding the controller as an additional node of the network, we
then are able to treat the pinning problem under the framework of
network synchronization \cite{MZ:2007}. In the enlarged network, the
controller has degree $n$, and is unidirectionally coupled to the
pinned nodes with strength $\eta$. So the enlarged system can be
regarded as a special example of weighted and directed network. The
enlarged network can be written as
\begin{displaymath}
W = \left( \begin{array}{ccccc}
a_{11} & a_{12} &  \ldots & a_{1N} & w_{1(N+1)} \\
\vdots &  \vdots & \ddots & \vdots & \vdots \\
a_{i1} &  a_{i2} &\ldots & a_{iN} & w_{i(N+1)}\\
\vdots &  \vdots & \ddots & \vdots & \vdots \\
a_{N1} &  a_{N2} & \ldots & a_{NN} &  w_{N(N+1)} \\
0 & 0& \ldots & 0 & 0
\end{array} \right),
\end{displaymath}
in which the controller is represented by the node of index $N+1$.
The pinnings are represented in the last column of $W$, where
$w_{i(N+1)}=\eta$ if $i\in V$, and $w_{i(N+1)}=0$ otherwise. Based
on the method of master-stability function \cite{MSF,MSF-HU}, we
know that whether the network can be ``synchronized" to the
controller is determined jointly by the node dynamics, $\mathbf{F}$,
and the eigenvalue spectrum of the coupling matrix $G=W-DI$. Here,
$D=(d_1,d_2,\ldots,d_{N+1})^T$ ($d_i=\sum_{j=1}^{N+1}w_{ij}$ is the
coupling intensity of node $i$) and $I$ is the identity matrix of
dimension $N+1$. Previous studies have shown that, for the general
types of node dynamics, the network is more synchronizable when the
spread of the eigenvalue spectrum is narrower
\cite{MSF,MSF-HU,HCLP:2009}. In particular, let
$0=\lambda_1>\lambda_2 \ldots \lambda_{N+1}$ be the eigenvalue
spectrum of the coupling matrix $G$. Then the smaller the eigenratio
$R=\lambda_{N+1}/\lambda_2$, the more likely synchronous dynamics is
to occur on the network. Since here we treat the pinning problem as
a special case of synchronization, smaller eigenratio thus also
means higher controllability.

\begin{figure}[tbp]
\begin{center}
\includegraphics[width=0.8\linewidth]{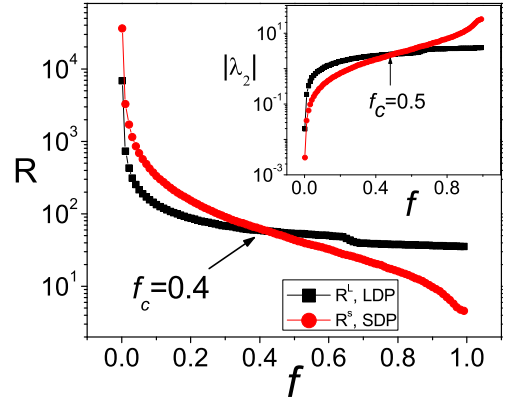}
\caption{(Color online) With pinning strength $\eta=25$, the
comparison of the performance between the LDP and SDP strategies in
controlling a scale-free network. Inset: the variations of the
largest non-zero eigenvalue as a function of the fraction of pinned
nodes. Each data is averaged over $20$ network realizations.}
\label{Fig_1}
\end{center}
\end{figure}

Having specified the pinning model, we now compare the performance
of the two pinning strategies, namely LDP and SDP, based on
numerical simulations. Without loss of generality, we generate a
complex network by the standard BA algorithm \cite{CN:REV}, which
consists of $N=1000$ nodes and has average degree $\left<
k\right>=8$. The node degree of the generated network follows
roughly a power-law distribution $p(k)\sim k^{-\gamma}$, with
$\gamma=3$. The LDP and SDP strategies are implemented as follows.
We first rearrange the nodes by a descending order of their degree,
i.e. $k_1 > k_2 \ldots> k_N$. Then, in LDP the $n$ largest-degree
nodes are pinned all together, i.e., nodes of index $1\leq m \leq n$
are pinned; while in SDP, the $n$ smallest-degree nodes are pinned
i.e., nodes of index $N-n+2 \leq m \leq N+1$ are pinned. The
eigenratio of the two pinned networks are denoted by $R^L$ (LDP) and
$R^S$ (SDP). The variations of $R^L$ and $R^S$ as a function of the
fraction of pinned nodes, $f=n/N$, are plotted in Fig. 1.
Interestingly, we find that, when the number of pinned nodes is
large enough, $R^S$ is smaller than $R^L$, i.e., the SDP strategy
has a higher performance than the LDP strategy. More specifically,
there exists a critical fraction, $f_c \approx 0.4$, in the number
of pinned nodes. When $f$ is small, as tacitly believed, the LDP
strategy has the higher performance than the SDP strategy. However,
as $f$ increases, the advantage of LDP over SDP is gradually
narrowed and, at the critical value $f_c$, the two strategies give
the same performance. After that, the SDP strategy will outperform
the LDP strategy. Moreover, as $f$ increases from $f_c$, the
advantage of SDP over LDP is gradually enlarged. The difference
between the two strategies reaches its maximum at $f=1-1/N$, where
only the small-degree (largest-degree) node is un-pinned in LDP
(SDP). Of course, at the point $f=1$ the network is globally
controlled, and the two strategies give the same performance again.
(The switch of control performance is also observed in the variation
of $\lambda_2$, which characterizes the system controllability for
the situation of non-bounded MSF function \cite{HCLP:2009}.)

\begin{figure}[tbp]
\begin{center}
\includegraphics[width=\linewidth]{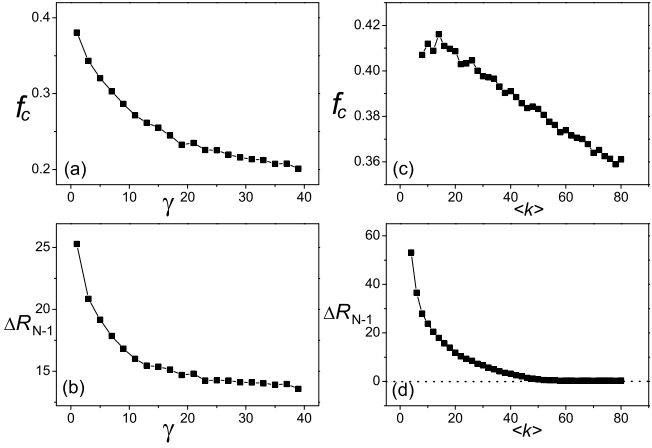}
\caption{The other parameters are the same as in Fig. 1. By varying
the degree exponent, the variations of (a) the critical fraction
$f_c$ and (b) the performance difference between SDP and LDP at
$n=N-1$. (b) and (c), the variations of $f_c$ and $\Delta R_{N-1}$
as a function of the network average degree. Each data is averaged
over $20$ network realizations.} \label{Fig_2}
\end{center}
\end{figure}

Is the above phenomenon general for complex networks? To answer this
question, we have compared the performance of the two strategies on
a variety of networks, including changing the degree exponent, the
average degree, the network size and the pinning strength of the
network model used in Fig. 1. In exploring the influence of the
degree exponent $\gamma$, we have adopted the network model of Ref.
\cite{DM:2002}, where $\gamma$ can be increased from $3$ to a large
value by tuning a parameter. In doing this, the network topology is
changing from heterogenous to homogeneous gradually. The variation
of the critical fraction $f_c$ as a function of $\gamma$ is plotted
in Fig. 2(a), where it is found that the value of $f_c$ is
monotonically decreased. Meanwhile, with the increase of $\gamma$,
the performance difference between the two strategies, $\Delta R =
R^{L}-R^{S}$, is narrowed, e.g, the variation of $\Delta R$ at
$n=N-1$ [Fig. 2(b)]. The narrowed difference in Fig. 2(b) seems to
implies that, comparing to homogeneous networks, heterogeneous
networks are more favorable for SDP. To check the influence of the
average degree, we have increased the value of $\left< k \right>$
from 6 to 80. The variations of the critical fraction and the
performance difference between the two strategies are plotted in
Fig. 2(c) and (d), respectively. It is seen that, as $\left< k
\right>$ increases, both $f_c$ and $\Delta R$ are decreased. Besides
degree exponent and average degree, we have also checked the
influence of the network size, and found that, as $N$ increases from
1000 to 5000, the values of $f_c$ and $\Delta R$ are decreased too
(not shown). Noticing the fact that either the increase of $\left< k
\right>$ or the decrease of $N$ leads to a homogeneous network
structure, these additional simulations therefore confirm the
finding of tuning $\gamma$, i.e., the SDP strategy for more
favorable in controlling heterogeneous networks.

The existence of $f_c$ can be mathematically analyzed, based on the
model of star networks. When not pinned, the star network has three
distinct eigenvalues: $\lambda_1=0$, $\lambda_N=-N$, and
$\lambda_i=-1$ for $i=2,\ldots,N-1$. When pinned, the eigenvalue
spectrum will be modified, reflected as the decreases of
$\lambda_1$, $\lambda_N$ and a few of other eigenvalues.
Specifically, when $n$ nodes are pinned, in LDP the number of
modified eigenvalues is $n+1$, while in SDP this number is $n+2$.
Firstly, let us consider the case of $n=1$ and compare the
performance of the two strategies. For LDP, only the central node is
pinned, and the two modified eigenvalues, $\lambda^L_1$ and
$\lambda^L_N$, are governed by equation
\begin{equation}
\lambda^2+(N+\eta)\lambda +\eta=0. \label{LDP}
\end{equation}
We thus have
$\lambda^L_{1,N}=[-(N+\eta)\pm\sqrt{(N+\eta)^2-4\eta}]/2$. For SDP,
the pinned node could be any of the $N-1$ peripheral nodes. The
three modified eigenvalues, $\lambda^S_1$, $\lambda^S_{N-1}$ and
$\lambda^S_N$, are governed by equation
\begin{equation}
-(\lambda^2+N\lambda+1)\eta=(\lambda+N)(\lambda+1)\lambda.
\label{SDP}
\end{equation}
By Cardan formula, we have $\lambda^S_1=-(N+1+\eta +\omega_1 p^{1/3}
+\omega_2 q^{1/3})/3$ and $\lambda^S_N=-(N+1+\eta +p^{1/3}
+q^{1/3})/3$, where $p=(a+ \sqrt{a^2-4b^3})/2$, $q=(a-
\sqrt{a^2-4b^3})/2$, $a=2(N+1+\eta)^3-9N(N+1+\eta)(1+\eta)+27\eta$,
$b=(N+1+\eta)^2-3N(1+\eta)$, and $\omega_{1,2}=-0.5\pm 0.5
\sqrt{3}i$. Previous studies have shown that, to control the
dynamics of a complex network successfully, the pinning strength
should be sufficiently large, i.e., $\eta>>1$
\cite{XFW:2002,XL:2004,TC:2007}. Under this condition, the above
eigenvalues are approximated as $\lambda^L_1 \approx -1$,
$\lambda^L_N \approx -(\eta+N)$, $\lambda^S_1 \approx -1/N$,
$\lambda^S_N \approx -(\eta+1)$. Apparently,
\begin{equation}
R^S=\lambda^S_N/\lambda^S_1>R^L=\lambda^L_N/\lambda^L_1,
\end{equation}
i.e., LDP has the higher performance than SDP when 1 node is pinned.

We next compare the performance of the two strategies for the
situation of $n=N-1$, i.e. at the end of the comparison. For LDP,
this means that except one peripheral node, all other nodes on the
network are pinned. In this case, the eigenvalues $\lambda^L_{N-1}$
and $\lambda^L_{N-1}$ are governed by equation
\begin{equation}
-[(N+2\lambda)\eta+\eta^2](\lambda+1)+\eta=(\lambda+N)(\lambda+1)\lambda.
\label{LDP2}
\end{equation}
While for SDP, only the central node is un-pinned. In this case, the
eigenvalues $\lambda^S_{N-1}$ and $\lambda^S_{N-1}$ are governed by
equation
\begin{equation}
\lambda^2+(N+\eta)\lambda+N\eta=0. \label{SDP2}
\end{equation}
Under the condition $\eta>>1$, these eigenvalues are approximated as
$\lambda^L_1 \approx -1$, $\lambda^L_N \approx -(\eta+1)$,
$\lambda^S_1 \approx -N$, $\lambda^S_N \approx -(\eta+1)$. So, when
$N-1$ nodes are pinned, we have
\begin{equation}
R^S=\lambda^S_N/\lambda^S_1<R^L=\lambda^L_N/\lambda^L_1,
\end{equation}
i.e., SDP has the higher performance than LDP. Since the situations
$n=1$ and $n=N-1$ stand as the two ends of the comparison [Fig. 1],
Eqs. [4] and [7] thus guarantee the existence of at least one
crossing in the performance curves.

The above phenomenon of performance switching can be physically
explained as follows. Still take the star-network as the model. When
only one node is pinned, by pinning the largest-degree node, the
pinning signal can be efficiently propagated to the peripheral
nodes, as the hub-node has the shortest average-node-distance. For
this reason, LDP will have a higher performance than SDP (see Eq.
[4]). However, this picture is changed when $N-1$ nodes are pinned.
Under a strong pinning strength ($\eta>>1$), in LDP the $N-1$ pinned
nodes have been well controlled to the target, and they will
influence the un-pinned peripheral node via a coupling of strength
$\varepsilon$; on the other hand, in SDP all the peripheral nodes
have been controlled, and they influence the central node together
via a joint coupling of strength $(N-1)\varepsilon$. Therefore in
SDP the coupling between the pinned part (the peripheral nodes) and
the un-pinned part is in fact amplified by $N-1$ times, therefore
making SDP superior to LDP. Having understood this mechanism, we are
able to predict that, as the size of the star network increases, the
advantage that SDP over LDP will be enlarged. Actually, this point
can be also drawn from Eq. [7], which implies that $\Delta R
=R^L-R^S \propto N$. As the degree of heterogeneity of star networks
can be characterized by the network size, this relation thus implies
from another angle the superiority of SDP in controlling
heterogeneous networks. It is worth noting that a similar phenomenon
about the role of small-degree nodes has been also observed in
network synchronization, where it is found that, to achieve global
network synchronization, it is usually the smallest-degree nodes
that are more important \cite{FMA:2006}.

\begin{figure}[tbp]
\begin{center}
\includegraphics[width=0.8\linewidth]{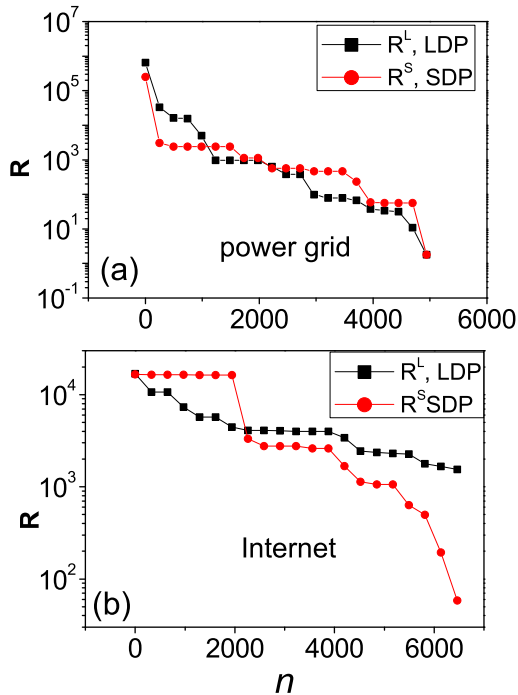}
\caption{(Color online) Comparison of performance the two pinning
strategies, LDP and SDP, for (a) the power-grid network of the
western United States and (b) the Internet at the Autonomous level.}
\label{Fig_3}
\end{center}
\end{figure}

How about the realistic networks? The specific examples we have
checked are the electrical power grid of the western United States
\cite{USPOWER} and the Internet at the autonomous level
\cite{INTERNET}. The power-grid network consists of $N=4941$ nodes
and has average degree $\langle k \rangle \approx 2.67$. The
variations of $R$ as a function of $n$ for both pinning strategies
are plotted in Fig. 3(a). Interestingly, it is found that, different
to the results of the ideal models [Fig. 1], the value of $R$ is
decreased in a non-smooth fashion and the two curves are crossed at
several places. We attribute this phenomenon to the inherent
community structures in the power-grid network (a detail analysis on
the non-smooth variation of $R$ will be presented elsewhere). The
Internet we have employed consists of $N=6474$ nodes and having an
average degree $\langle k \rangle \approx 3.88$. The variations of
$R$ as a function of $n$ for both pinning strategies are plotted in
Fig. 3(b). Again, we find a non-smooth decrease of $R$. For the
Internet network, we find only one crossing between the two curves,
which is located at about $n=2236$.

The finding that heterogenous networks can be better controlled by
pinning the small-degree nodes might have some potential
applications, say, for example, in synchronizing sensor networks or
controlling the consensus of multi-agent systems. In a sensor
network, due to terrain complexity the sensors are distributed in a
non-uniform fashion: most of the sensors have very few neighbors,
but a few sensors could have many neighbors \cite{SENSOR}. That is,
the sensor network has a heterogenous degree distribution. To
synchronize the timers of the sensors, a general approach is to
assemble a fraction of the sensors with a signal receiver, which
receives signal from the satellite and reset the timer of the
sensors accordingly. For those sensors without the receiver, their
timers are updated by their neighbors. Here an important question is
that, for the given number of sensors having signal receiver, how to
place them properly in order of a higher performance for timer
synchronization? The study of the current paper suggests that, if
the number of receiver-assembled sensors are large enough, it may be
better to place them in the areas of sparse sensors. Another
application could be the control of consensus in multi-agent systems
\cite{LS:2009}. To make the mobile agents move in a synchronous
fashion, in engineering one practical approach is to introduce some
``leading" agents into the system. Different to the normal agents,
these leading agents have a predefined motion, but they will
influence the motion of the normal agents via couplings (behave like
the pinning controller). Our study suggests that, to control the
multi-agent system more efficiently, we may choose to put the
``leading" agents into the small swarms, instead of the tacitly
believed large swarms. (From the network point of view, large and
small swarms could be regarded as possessing large and small degree,
respectively.)

In summary, we have compared the performance of two opposite pinning
strategies in controlling complex networks. It is found that, when a
significant fraction of the network nodes are pinned, the pinning of
the small-degree nodes could have a higher performance than the
large-degree ones. We have explained this phenomenon by a model of
simplified network, and discussed the dependence of this phenomenon
to the network parameters in detail. These findings, hopefully, will
provide a new angle to the pinning control of complex networks, and
apply to practical situations.

This work was supported by National Natural Science Foundation of
China under Grant No. 10805038 and by Chinese Universities
Scientific Fund.

\end{document}